\documentstyle[preprint,aps]{revtex}

\begin{document}
\draft
\setcounter{footnote}{1}
\title{Discrete Symmetries and Transformations of the Hubbard Model.} 
\author{J.P.Wallington and James F.Annett.}
\address{University of Bristol, H.H. Wills Physics Laboratory, 
Royal Fort, Tyndall Ave,
Bristol BS8 1TL, United Kingdom.}
\date{\today}
\maketitle

\begin{abstract}
We show that, in addition to $SO(4)$, the Hubbard model at half
filling on a bipartite lattice has a group of discrete symmetries and
transformations.  A unique Hubbard-Stratonovich decomposition of the
interaction term, incorporating both spin and pseudospin variables on
an equal footing, is found in which these symmetries are manifestly
present.  The consequences of this at the mean field and one loop
renormalisation group levels are discussed.
\end{abstract}
The Hubbard model is a simple, generic model of correlated electrons
on a lattice. It is defined by the Hamiltonian:
\begin{eqnarray}
&\hat{H}  & = \hat{H}_0 + \hat{H}_I - \mu'\sum_{r,\sigma}n_{\sigma}(r), \\
&\hat{H}_0& =
 -\sum_{\langle rr' \rangle,\sigma}t_{rr'}
              \left( c_{\sigma}^{\dagger}(r)  c_{\sigma}(r') +
		     h.c. \right), \\
&\hat{H}_I& = U \sum_{r} \left[n_{\uparrow}(r)-\frac{1}{2}\right]\left[n_{\downarrow}(r)-\frac{1}{2}\right],
\end{eqnarray}
where $t_{rr'}$ connects only nearest-neighbour lattice sites and
$\mu'=\mu-\frac{U}{2}$ controls the number density of the system. $U$
is the interaction energy.

The model was introduced by Hubbard \cite{Hubbard63} as a model of
itinerant ferromagnets.  Since then the repulsive case $(U>0)$ has
been used to describe both ferro- and antiferro-magnetism
\cite{Nagaoka,Hirsch} and $d$-wave superconductivity
\cite{Rice,Scalapino}.  For an attractive interaction $(U<0)$ it
has been applied to the study of BCS $s$-wave superconductivity
\cite{Scalettar} and the transition to the Bose condensation
of strongly bound pairs \cite{Micnas}.  Interest in the Hubbard model
on a 2D square lattice has been particularly intense since the
suggestion by Anderson \cite{Anderson} that it is relevant to the
description of the high-$T_c$ compounds.

Despite this interest the Hubbard model remains unsolved except in 1D.
However, in recent years, there have been a number of exact results
proved. Yang and Zhang \cite{SO4} have
shown, on a bipartite lattice at half-filling and zero field, the
existence of an $SO(4)
\equiv SU_S(2)\otimes SU_R(2)/Z_2$ symmetry.  The group's action
generates rotations of the spin and `pseudospin' vectors, ${\bf S}$
and ${\bf R}$:
\begin{eqnarray}
{\bf S} &\equiv& 
    \frac{1}{2}\Psi_S^{\dagger}(r){\mathbf \sigma}\Psi_S(r), \quad\quad
    \Psi_S(r) = \left(  \begin{array}{c}
        c_{\uparrow}(r) \\
        c_{\downarrow}(r)
                        \end{array}
                \right),\\
{\bf R} &\equiv& 
    \frac{1}{2}\Psi_R^{\dagger}(r){\mathbf \sigma}\Psi_R(r), \quad\quad
    \Psi_R(r) = \left(\begin{array}{c}
        c_{\uparrow}(r) \\
        e^{\imath\pi.r} c_{\downarrow}^{\dagger}(r)
                      \end{array}
                \right),
\end{eqnarray}
where ${\mathbf \sigma}$ is the vector of Pauli matrices and $\pi$ is
the vector $(\pi,\pi,...)$ in $d$ dimensions.  The pseudospin vector
comprises superconducting pair creation/annihilation operators and the
number operator.  A consequence of pseudospin rotational invariance is
the existence, at half filling, of a degenerate
superconducting/charge density wave (S-CDW) ordering
\cite{Gyorffy,Wallington}. {\"O}stlund \cite{Ostlund} has given a
thorough account of the $SO(4)$ symmetry, including an enumeration of
the possible mean field phases of the model.

Recently Zhang \cite{ZhangSO5} proposed an approximate
$SO(5)$ symmetry connecting antiferromagnetism to $d$-wave
superconductivity.  The proposal offers the possibility of
understanding the phase diagram of the cuprates and has attracted much
attention.  The symmetry is not exact, unlike $SO(4)$, but becomes valid
under the action of the renormalisation group. 

In this work we examine exact symmetries.  We show that in addition to
the continuous $SO(4)$ there are non-trivial discrete symmetries.
These play a similar role to CPT invariance in relativistic quantum
mechanics.  In fact, this analogy is very close, as will be clear
below, because $SO(4)$ is the Euclidean analogue of the Lorentz group,
$SO(3,1)$. We show that these symmetries, augmented by the addition of
the Lieb-Mattis transformation, are very useful in resolving an
ambiguity inherent in any field theoretic treatment based upon the
Hubbard-Stratonovich transformation \cite{Hub_HST,Strat_HST}:
\begin{equation}
e^{\frac{1}{2}{\cal F}^2} \equiv \int{\cal D} \left[ b \right]
e^{-\frac{1}{2}b^2-{\cal F}b},
\label{eqn:HST}
\end{equation}
where ${\cal F}$ is a bilinear Fermi operator and $b$ is a Bose field.
The use of this identity converts the problem of interacting fermions
into one of free fermions interacting with bosons. To employ this
technique the quartic term, $\hat{H}_I$, must be written as the square
of a bilinear operator. The method is ambiguous because there are many
ways (actually a continuous infinity) of doing this \cite{Schumann}:
\begin{equation}
\label{eqn:decomp_alpha}
\hat{H}_I = \sum_{r}\left[
            -\frac{U}{3}\left(1-\alpha \right){\bf S}\cdot{\bf S}
            +\frac{U}{3}\left(1+\alpha \right){\bf R}\cdot{\bf R}\right].
\end{equation}
Whilst this is formally exact for all $\alpha$, each choice behaves
differently under approximation schemes (e.g. mean field theory, RPA,
renormalisation group).  It is customary to pick a decomposition which
is most convenient for the problem at hand; for instance $\alpha=-1$
for magnetism or $\alpha=+1$ for superconductivity. We will show,
however, using the discrete transformation properties of the
Hamiltonian, that there is a single `best' decomposition.  Uniquely,
approximation techniques applied to this decomposition preserve all
the symmetry and transformation properties of the full Hamiltonian.

It is convenient to introduce a four spinor:
\begin{equation}
\Psi(r)\equiv\left(c_{\uparrow}(r),c_{\downarrow}(r),e^{\imath\pi.r}
c_{\uparrow}^{\dagger}(r),e^{\imath\pi.r}
c_{\downarrow}^{\dagger}(r)\right)^T,
\end{equation}
in terms of which both ${\bf S}$ and ${\bf R}$ may be
constructed. Equivalently a $2\times2$ matrix form is possible \cite{Schulz}.

We consider three discrete symmetries; charge conjugation,
$\hat{{\cal C}}$, parity, $\hat{{\cal P}}$, and time reversal, $\hat{{\cal
T}}$, also the Lieb-Mattis canonical transformation, $\hat{Z}$.

{\em Charge Conjugation}, $\hat{{\cal C}}$. At half filling ($\mu'=0$)
on a bipartite lattice, for every state of energy $\epsilon({\bf k})$
below the Fermi energy there is a corresponding state of energy
$\epsilon({\bf k}+{\mathbf\pi})=-\epsilon({\bf k})$ above it. Under
charge conjugation a hole of momentum ${\bf k}$ is equivalent to a
particle of momentum ${\bf k}+{\mathbf \pi}$. Interpreting filled states
as the Dirac sea and holes as positrons, the situation is analogous to
the relativistic case. With respect to the four spinor, $\Psi(r)$, charge
conjugation is defined by:
\begin{equation}
\hat{{\cal C}} \equiv \left(
                      \begin{array}{cccc}
                      0 &  0 & 0 & -1 \\
                      0 &  0 & 1 &  0 \\
                      0 & -1 & 0 &  0 \\
                      1 &  0 & 0 &  0 
                      \end{array}
                      \right) \otimes \hat{Y},
\end{equation}
where $\hat{Y}$ is the antilinear complex conjugation operator. Its
effects on the spin and pseudospin operators are as follows:
\begin{eqnarray}
\hat{{\cal C}}{\bf R}\hat{{\cal C}}^{-1} & = & -{\bf R},\\
\hat{{\cal C}}S^{\pm}\hat{{\cal C}}^{-1} & = & S^{\pm},\\
\hat{{\cal C}}S_z\hat{{\cal C}}^{-1} & = & S_z.
\end{eqnarray}

{\em Parity}, $\hat{{\cal P}}$, leaves ${\bf S}$, ${\bf
R}$ and the Hamiltonian unchanged and is of little use to us here.
 
{\em Time Reversal}, $\hat{{\cal T}}$, reverses the signs of the
linear and angular momenta:
\begin{equation}
\hat{{\cal T}}\hat{{\bf P}}\hat{{\cal T}}^{-1}  =  -\hat{{\bf P}},\quad
\hat{{\cal T}}{\bf S}\hat{{\cal T}}^{-1}  =  -{\bf S}.
\end{equation}
Cooper pairs are time reversal invariant as is charge, thus:
\begin{equation}
\hat{{\cal T}}R^{\pm}\hat{{\cal T}}^{-1} =  R^{\pm},\quad
\hat{{\cal T}}R_z\hat{{\cal T}}^{-1} =  R_z.
\end{equation}
Using these conditions, we may deduce the form of the time reversal
operator acting on $\Psi(r)$:
\begin{equation}
\hat{{\cal T}} \equiv       \left(
                            \begin{array}{cccc}
                             0 & -1 & 0 &  0 \\
                             1 &  0 & 0 &  0 \\
                             0 &  0 & 0 & -1 \\
                             0 &  0 & 1 &  0 
                            \end{array}
                            \right) \otimes \hat{Y}. 
\end{equation}
The antilinear operator, $\hat{Y}$, is necessary to ensure that the
commutation relations,
$\hat{P}_{\alpha}\hat{Q}_{\alpha}-\hat{Q}_{\alpha}\hat{P}_{\alpha}=i\hbar$
are preserved under $\hat{{\cal T}}$.

The {\em Lieb-Mattis canonical transformation}, $\hat{Z}$, is defined
by: $c_{\uparrow}(r) \mapsto c_{\uparrow}(r)$, $c_{\downarrow}(r)
\longmapsto e^{\imath{\mathbf\pi}\cdot{\bf
r}}c_{\downarrow}^{\dagger}(r)$.  It acts on the four-spinor as a
matrix:
\begin{equation}
\Psi(r) \stackrel{\hat{Z}}{\longmapsto} 
       \left(
          \begin{array}{cccc}
                             1 & 0 & 0 & 0 \\
                             0 & 0 & 0 & 1 \\
                             0 & 0 & 1 & 0 \\
                             0 & 1 & 0 & 0 
          \end{array}
    \right)\Psi(r).
\end{equation}
It maps spin into pseudospin and vice versa, leaves the
non-interacting Hamiltonian, $\hat{H}_0$, unchanged but reverses the
sign of the interaction term, $\hat{H}_I$:
\begin{eqnarray}
{\bf S} & \stackrel{\hat{Z}}{\Longleftrightarrow} & {\bf R},\\
\hat{H}_0 + \hat{H}_I & \stackrel{\hat{Z}}{\Longleftrightarrow} & 
  \hat{H}_0 - \hat{H}_I.
\end{eqnarray}
The action of $\hat{Z}$ on the operators $\hat{\cal C}$ and $\hat{\cal
T}$ is noteworthy:
\begin{equation}
\hat{\cal C} = \hat{Z}\hat{\cal T}\hat{Z},
\end{equation}
implying that the behaviour of {\bf S} under $\hat{\cal T}$ is the
same as {\bf R} under $\hat{\cal C}$.

{\em The Discrete Group of Transformations}. Using the matrix forms of
$\hat{\cal C}$, $\hat{\cal T}$ and $\hat{Z}$ we see that the set,
${\cal G} \equiv \{\pm{\bf I},$ $\pm{\hat{\cal C}},$ $\pm{\hat{\cal
T}},$ $\pm{\hat{Z}},$ $\pm{\hat{\cal C}}{\hat{\cal T}},$
$\pm{\hat{Z}}{\hat{\cal C}},$ $\pm{\hat{Z}}{\hat{\cal T}},$
$\pm{\hat{\cal C}}{\hat{\cal T}}{\hat{Z}}\}$, forms a discrete group
under multiplication.  It comprises 16 elements, falling into ten
conjugacy classes. The character table is shown in Table
\ref{table:characters}.

Each of the operators $\hat{H}_0$, $\hat{H}_I$, ${\bf S}$ and ${\bf
R}$ may be classified as transforming according to a certain
representation: $\hat{H}_0$ transforms as the identity representation,
$\Gamma^{(1)}$, $\hat{H}_I$ transforms as $\Gamma^{(2)}$ whilst ${\bf
S}$ and ${\bf R}$ transform as couplets;
 ${S_x \choose R_x}$, ${S_z
\choose R_z}$ as $\Gamma^{(5)}$ and ${S_y \choose R_y}$ as
$\Gamma^{(3)}\oplus\Gamma^{(4)}$.

The space of bilinear operators is spanned by ${\bf S}$ and ${\bf
R}$. Therefore the quartic operator, $\hat{H}_I$, can be
decomposed into products of ${\bf S}$ and ${\bf R}$.  The
representations of all possible products are generated by:
\begin{eqnarray}
&&\left(\Gamma^{(3)}\oplus\Gamma^{(4)}\oplus2\Gamma^{(5)}\right)
\otimes\left(\Gamma^{(3)}\oplus\Gamma^{(4)}\oplus2\Gamma^{(5)}\right)\\
&&\quad=\quad 6\Gamma^{(1)}\oplus6\Gamma^{(2)}\oplus4\Gamma^{(3)}\oplus4\Gamma^{(4)}\oplus8\Gamma^{(5)}\nonumber.
\end{eqnarray}
Discarding all but those products which transform as $\Gamma^{(2)}$
and imposing explicit spin and pseudospin invariance we arrive at the
unique decomposition:
\begin{equation}
\label{eqn:decomp}
\hat{H}_I=\sum_{r}\left[-\frac{U}{3}{\bf S}\cdot{\bf S}+\frac{U}{3}{\bf R}\cdot{\bf R}\right].
\end{equation}
We have constructed this decomposition (corresponding to $\alpha=0$ in
equation \ref{eqn:decomp_alpha}) in such a way that all its symmetries
are transparent and are inherited from the transformation properties
of ${\bf S}$ and ${\bf R}$.  Because of this it enjoys a great
advantage over the other choices of $\alpha$. This becomes clear when we
consider approximation schemes.

{\em Mean Field Theory}. Using equation \ref{eqn:decomp_alpha} we
construct a mean field theory for each $\alpha$:
\begin{eqnarray}
\label{eqn:decomp_mft}
\hat{H}_I^{MF} &=& \sum_{r}\left[-\frac{U}{3}\left(1-\alpha \right)\left\{2{\bf S}\cdot\left<{\bf S}\right>-\left<{\bf S}\right>\cdot\left<{\bf S}\right>\right\}\right.\nonumber\\
&+&\left.\frac{U}{3}\left(1+\alpha \right)\left\{2{\bf R}\cdot\left<{\bf R}\right>-\left<{\bf R}\right>\cdot\left<{\bf R}\right>\right\}\right].
\end{eqnarray}
The expectations, $\left<...\right>$, are taken self consistently with
respect to the mean field Hamiltonian. For all $\alpha$ and for $U>0$,
the solutions are either paramagnetic ($T>T_c$) or antiferromagnetic
($T<T_c$) but in both cases $\left<{\bf R}\right>=0$.  Similarly for
$U<0$, there is a S-CDW phase for $T<T_c$ and $\left<{\bf S}\right>=0$.

The full theory is invariant under the combined action of $\hat{Z}$
and $U\mapsto-U$, therefore we expect a good mean field theory to
share this property.  For instance, the values of $\left<{\bf
S}\right>$ (for $U>0$) and $\left<{\bf R}\right>$ (for $U<0$) should
not change.  We find that this is true for the $\alpha=0$ theory but
not for any other.  This can be simply understood by considering the
action of $\hat{Z}$ and $U\mapsto-U$ on equation \ref{eqn:decomp_mft}.
The effective couplings which the fields `see',
$\frac{U}{3}\left(1-\alpha \right)$ for {\bf S} and
$\frac{U}{3}\left(1+\alpha \right)$ for {\bf R}, are exchanged, the
effect of which is to alter the expectation values $\left<{\bf
S}\right>$ and $\left<{\bf R}\right>$.  This has dramatic consequences
when $\alpha=\pm1$ in which cases the fields of interest acquire zero
couplings and drop out of the Hamiltonian altogether.  The $\alpha=0$
theory avoids these problems by having the same coupling,
$\frac{U}{3}$, for both {\bf S} and {\bf R}.

It is desirable that Hartree-Fock theory be reproduced by the scalar
version of equation \ref{eqn:decomp_alpha}:
\begin{equation}
\hat{H}_I = \sum_{r}\left[
            -U\left(1-\alpha \right)S_z^2
            +U\left(1+\alpha \right)R_z^2\right],
\end{equation}
where $S_z=\frac{1}{2}(n_{\uparrow}-n_{\downarrow})$ and $R_z
=\frac{1}{2}(n_{\uparrow}+n_{\downarrow}-1)$. Hartree-Fock is only
recovered for the $\alpha=0$ decomposition.  For $\alpha \neq 0$ there
are extra terms, $n_{\sigma}\left<n_{\sigma}\right>$, which violate
the Pauli exclusion principle.  These arise because mean field theory
does not respect idempotency, i.e. $n_{\sigma}n_{\sigma}= n_{\sigma}$
but $n_{\sigma}\left<n_{\sigma}\right>\neq n_{\sigma}$.  The
$\alpha=0$ theory was derived without the use of idempotency and so
reproduces Hartree-Fock theory automatically.

{\em Landau-Ginzburg Theory}. Because the $\alpha=0$ decomposition
retains both {\bf S} and {\bf R} it has a further advantage in that,
combined with the use of $\Psi(r)$, it allows us to study the
renormalising effects of magnetism on superconductivity and vice
versa.  To see these we construct a Landau-Ginzburg (LG) functional.

Using equation (\ref{eqn:decomp}) in the Hubbard Stratonovich
transformation and then integrating out the fermionic degrees of
freedom leaves an effective bosonic action in terms of the fields
$\Phi$ and $\Delta$ which couple to ${\bf S}$ and ${\bf R}$.  In the
static limit the quadratic part reads:
\begin{eqnarray}
{\cal S}^{(2)}&=&
\frac{1}{2}\sum_q\left(1-\frac{U}{3}\chi(q)\right)\Phi(q)\cdot\Phi(-q)\\
&+&\quad\frac{1}{2}\sum_q\left(1+\frac{U}{3}\chi(q)\right)\Delta(q)\cdot\Delta(-q),
\nonumber
\end{eqnarray}
where the susceptibility is:
\begin{equation}
\chi(q)=-\sum_k \frac{f(\epsilon_{\bf k+q})-f(\epsilon_{\bf k})}{\epsilon_{\bf k+q}-\epsilon_{\bf k}}.
\end{equation}
To construct the LG functional we need only long wavelength (small q)
excitations about stable modes.  The sign difference in the
expressions for $\Phi$ and $\Delta$ means that (for $U>0$) the stable
modes are at $\pi$ (antiferromagnetism) and 0 ($\eta$-pairing \cite{Yang}) for
$\Phi$ and $\Delta$ respectively. For $U<0$ the stable modes of $\Phi$
and $\Delta$ are at 0 (ferromagnetism) and $\pi$
(superconductivity/CDW). Taking $U>0$ and expanding in q about the
stable momenta gives the LG form:
\begin{eqnarray}
\label{eqn:lg}
{\cal S}&=&
\frac{1}{2}\sum_q\left(r_{\pi}+\frac{q^2}{2m_{\pi}}\right)\Phi(q)\cdot\Phi(-q)\\
&+&\quad
\frac{1}{2}\sum_q\left(r_0+\frac{q^2}{2m_0}\right)\Delta(q)\cdot\Delta(-q)\nonumber\\
&+&\quad
u_{\pi}\sum_{q_1,q_2,q_3}
\left(\Phi_1\cdot\Phi_2\right)
\left(\Phi_3\cdot\Phi_{-1-2-3}\right)\nonumber\\
&+&\quad 
u_0\sum_{q_1,q_2,q_3}
\left(\Delta_1\cdot\Delta_2\right)
\left(\Delta_3\cdot\Delta_{-1-2-3}\right)\nonumber\\
&+&\quad
u_x\sum_{q_1,q_2,q_3}
\left(\Delta_1\cdot\Delta_2\right)
\left(\Phi_3\cdot\Phi_{-1-2-3}\right)\nonumber,
\end{eqnarray}
where $\Phi_i\equiv\Phi(q_i)$ (similarly for $\Delta$) and:
\begin{eqnarray}
r_{\pi}  &=& 1-\frac{U}{3}\chi({\pi}),\quad\quad 
r_0 = 1+\frac{U}{3}\chi(0),\\
\frac{1}{m_{\pi}} &=& -\frac{U}{3}\frac{\partial^2\chi({\pi})}{\partial q^2}
,\quad\quad 
\frac{1}{m_0} = \frac{U}{3}\frac{\partial^2\chi(0)}{\partial q^2},\\
u_{\pi} &=& \frac{U^2}{72}\sum_{\epsilon}\left[
\frac{f'(\epsilon)}{\epsilon^2}-\frac{f(\epsilon)}{\epsilon^3}+\frac{1}{2\epsilon^3}\right]N(\epsilon),\\
u_0     &=& \frac{U^2}{216}\sum_{\epsilon}f'''(\epsilon)N(\epsilon),\\
 u_x &=& \frac{U^2}{72}\sum_{\epsilon}\frac{f''(\epsilon)}{\epsilon}N(\epsilon),
\end{eqnarray}
where $N(\epsilon)$ is the non-interacting density of states.  For
$U<0$ the roles of $\Phi$ and $\Delta$ are simply reversed.

Equation \ref{eqn:lg} is the starting point for a renormalisation
group (RG) analysis. We employ a one loop technique {\em \`{a} la}
Wilson-Kogut
\cite{RG1,RG2}; i.e. successive elimination of `fast-modes' (large q)
followed by rescaling of fields and momenta.  This process
renormalises the parameters $r_{\pi,0}$ and $u_{\pi,0,x}$. The
analysis shows that the cross term, $u_x$, is irrelevant, i.e. it
flows to zero under the action of the RG. Therefore at the fixed point
the spin and pseudospin are decoupled from each other.  This implies
that for $U>0$ the fixed point is the usual antiferromagnetic one,
whilst it is S-CDW for $U<0$.  The sole effect of $u_x$ is to
renormalise $T_c$.  Fig.~\ref{figure1} shows RG flows for different initial
$u_x$ but the same temperature, $T_c(u_x=0)$. For $u_x>0$ the flow
goes to the high temperature phase implying
$T_c(u_x)<T_c(0)$. Conversely for $u_x<0$ the ordered phase (AFM or
S-CDW) is reached implying $T_c(u_x)>T_c(0)$. To conclude, $T_c$ is
depressed for positive $u_x$ but increased for negative $u_x$.

In summary, we have discovered a discrete group of symmetries and
transformations of the Hubbard model on a bipartite lattice at half
filling and enumerated its representations.  Using these we have
constructed a unique Hubbard Stratonovich decomposition of the Hubbard
interaction term, which treats spin and pseudospin on an equal
footing. At a mean field level it is superior to other
decompositions because it preserves all the
transformation properties of the full theory.  In addition, the unique
decomposition we have derived has allowed us to construct a
Landau-Ginzburg functional to examine the interplay of magnetism and
superconductivity/CDW. For $U>0$, the critical temperature for the
onset of antiferromagnetism, $T_c^{AFM}$, is found to be renormalised
by $\eta$-paired fluctuations.  Reciprocally, for $U<0$, $T_c^{S-CDW}$
is renormalised by ferromagnetic fluctuations.  Critical exponents are
not affected.

This work is supported by the Engineering and Physical Sciences
Research Council (EPSRC) through the provision of a studentship to
J.P.W. and by the Office of Naval Research Grant No. N00014-95-1-0398
(J.F.A.).

\begin{figure}
\caption{RG flows for different initial values of $u_x$. For $u_x=0$ the full
lines show flows starting infinitessimally above and below $T_c$.  The
broken lines are for the same temperature but non-zero initial $u_x$.}
\label{figure1}
\end{figure}

\widetext
\begin{table}
\caption{\label{table:characters}Character Table for the group of $\hat{\cal C}$, $\hat{\cal T}$ and $\hat{Z}$.}
\begin{tabular}{|l||c|c|c|c|c|c|c|c|c|c|}\hline
  &\makebox[1.4cm]{} &\makebox[1.2cm]{} &\makebox[1.2cm]{}
  &\makebox[1.2cm]{} &\makebox[1.2cm]{$\hat{\cal C}$}
  &\makebox[1.2cm]{-$\hat{\cal C}$} &$\hat{Z}$ &-$\hat{Z}$
  &\makebox[1.2cm]{$\hat{\cal C}\hat{Z}$} &\makebox[1.2cm]{-$\hat{\cal
  C}\hat{Z}$} \\ &\raisebox{1.5ex}[0pt]{\bf I}
  &\raisebox{1.5ex}[0pt]{-\bf I} &\raisebox{1.5ex}[0pt]{${\hat{\cal
  C}}{\hat{\cal T}}$} &\raisebox{1.5ex}[0pt]{-${\hat{\cal
  C}}{\hat{\cal T}}$} &${\hat{\cal T}}$ & -${\hat{\cal T}}$
  &\makebox[1.2cm]{-$\hat{\cal C}\hat{\cal T}\hat{Z}$}
  &\makebox[1.2cm]{${\hat{\cal C}}{\hat{\cal T}}{\hat{Z}}$}
  &${\hat{\cal T}}{\hat{Z}}$ &-${\hat{\cal T}}{\hat{Z}}$\\
\hline\hline
$\Gamma^{(1)}$ & 1 & 1 & 1 & 1 & 1 & 1 & 1 & 1 & 1 & 1\\
\hline
$\Gamma^{(2)}$ & 1 & 1 & 1 & 1 & 1 & 1 & -1 & -1 & -1 & -1\\
\hline
$\Gamma^{(3)}$ & 1 & 1 & 1 & 1 & -1 & -1 & -1 & -1 & 1 & 1\\
\hline
$\Gamma^{(4)}$ & 1 & 1 & 1 & 1 & -1 & -1 & 1 & 1 & -1 & -1\\
\hline
$\Gamma^{(5)}$ & 2 & 2 & -2 & -2 & 0 & 0 & 0 & 0 & 0 & 0\\
\hline
$\Gamma^{(6)}$ & 1 & -1 & -1 & 1 & i & -i & 1 & -1 & i & -i\\
\hline
$\Gamma^{(7)}$ & 1 & -1 & -1 & 1 & i & -i & -1 & 1 & -i & i\\
\hline
$\Gamma^{(8)}$ & 1 & -1 & -1 & 1 & -i & i & -1 & 1 & i & -i\\
\hline
$\Gamma^{(9)}$ & 1 & -1 & -1 & 1 & -i & i & 1 & -1 & -i & i\\
\hline
$\Gamma^{(10)}$ & 2 & -2 & 2 & -2 & 0 & 0 & 0 & 0 & 0 & 0\\
\hline
\end{tabular}
\end{table}
\narrowtext

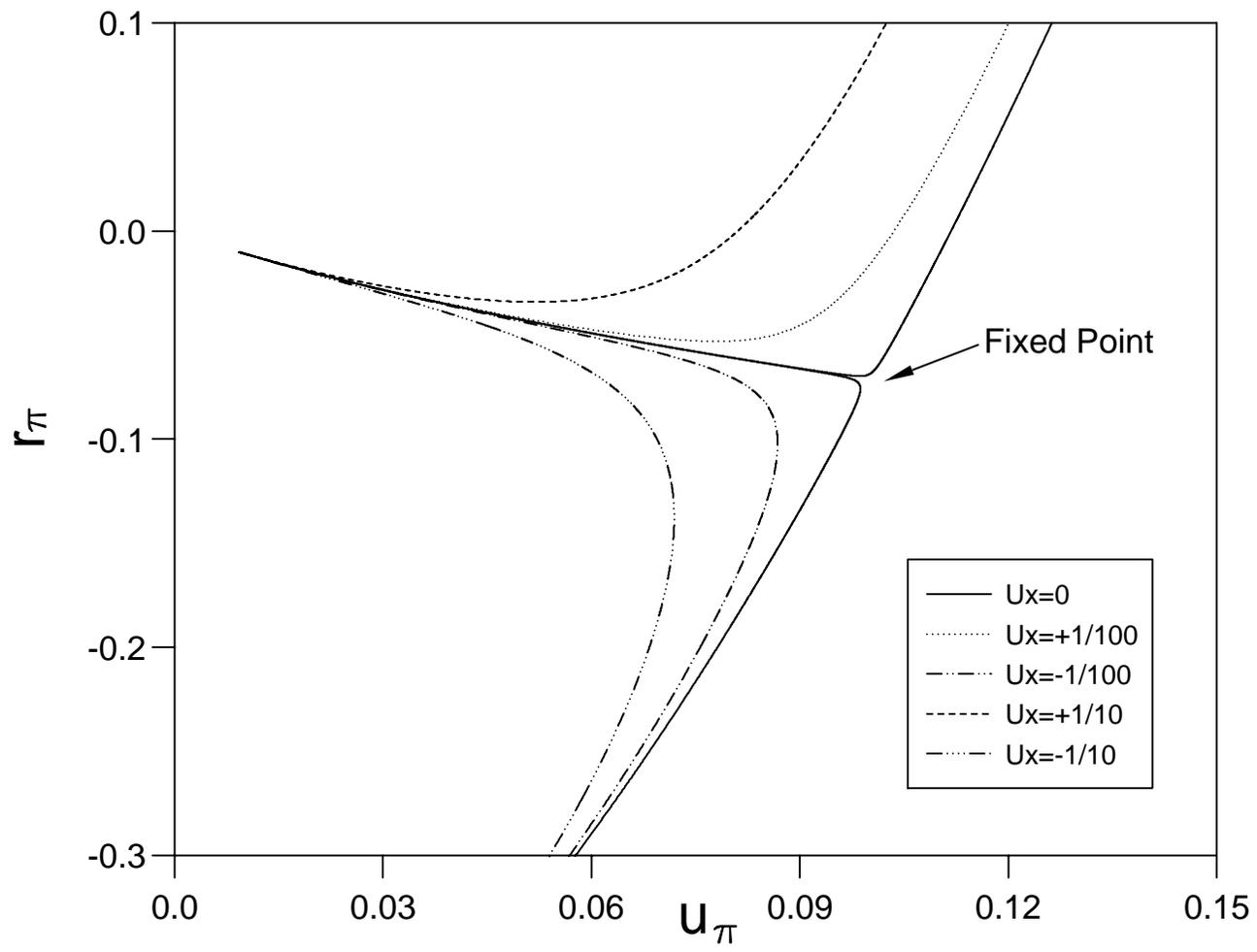
\end{document}